\documentclass[preprint,review,12pt]{elsarticle}

\usepackage{graphicx}
\usepackage{amsfonts}
\usepackage{amsmath}
\usepackage{amssymb}
\usepackage{color}
\usepackage{array}
\usepackage{xcolor}
\usepackage{soul}

\newcounter{bla}

\journal{Computer Physics Communications}

\newcommand{\sug}[1]{#1}

\begin{document}

\begin{frontmatter}

\title{A new method to dispatch split particles in Particle-In-Cell codes}

\author[a]{Roch Smets\corref{author}}
\author[a]{Nicolas Aunai}
\author[b]{Andrea Ciardi}
\author[a,b]{Matthieu Drouin}
\author[c]{Martin Campos-Pinto}
\author[a]{Philip Deegan}

\cortext[author] {Corresponding author.\\\textit{E-mail address:} roch.smets@upmc.fr}
\address[a]{Sorbonne Universit\'e, LPP, CNRS UMR 7648, Ecole Polytechnique, Universit\'e Paris Sud, Observatoire de Paris, Paris, France}
\address[b]{Sorbonne Universit\'e, Observatoire de Paris, PSL Research University, LERMA, CNRS UMR 8112, Paris, France}
\address[c]{Sorbonne Universit\'e, LJLL, CNRS UMR 7598, Paris, France}

\begin{abstract}
Particle-In-Cell codes are widely used for plasma physics simulations. It is often the case that particles within a computational cell need to be split to improve the statistics or, in the case of non-uniform meshes, to avoid the development of fictitious self-forces. Existing particle splitting methods are largely empirical and their accuracy in preserving the distribution function has not been evaluated in a quantitative way. Here we present a new method specifically designed for codes using adaptive mesh refinement. Although we point out that an exact, distribution function preserving method does exist, it requires a large number of split particles and its practical use is limited. We derive instead a method that minimizes the cost function representing the distance between the assignment function of the original particle and that of the sum of split particles. Depending on the interpolation degree and on the dimension of the problem, we provide tabulated results for the weight and position of the split particles. This strategy represents no overhead in computing time and for a large enough number of split-particles it asymptotically tends to the exact solution.
\end{abstract}


\end{frontmatter}

\begin{small}
\noindent
{\em Nature of problem(approx. 50-250 words):}\\
The macro-particles in an AMR PIC code need to be split when traveling from a coarse region to a finer one. No mathematically rigorous way of doing so has yet been proposed. Specifically, splitting can lead to the enhancement at an unacceptable level of the fluctuation level of the particles moments (density, current...).\\
{\em Solution method(approx. 50-250 words):}\\
We propose a deterministic method based on the minimization of the difference between the assignment function of the particle to be split and the one of the set of particles resulting from the splitting.\\
{\em Additional comments including Restrictions and Unusual features (approx. 50-250 words):}\\
\end{small}

\section{Introduction}

The multi-scale character of laboratory and astrophysical plasmas is ubiquitous. Examples include, but are not limited to, the turbulence in the solar wind or in fusion experiments\cite{parashar2010}, the reconnection of magnetic field lines\cite{kulsrud2011}, the formation and propagation of shocks\cite{giacalone2007}. In all these systems, spatial and temporal scales can span many orders of magnitude and impose stringent constraints on grid-based numerical codes. In situations where small scales are spatially localized, a solution to circumvent some of the constraints is to adaptively refine the mesh in the regions of interest. These Adaptive Mesh Refinement (AMR) technique are widely used in fluid codes (see e.g. \cite{berger1989}, \cite{teyssier2002}) and have also begun to be successfully implemented in Particle-In-Cell (PIC) codes (\sug{\cite{vay2004}}, \cite{fujimoto2006}, \cite{muller2011}, \sug{\cite{usui2011}}, \cite{innocenti2013}, \sug{\cite{kolobov2016}, \sug{\cite{markidis2018}}}).

In PIC codes, a large collection of physical particles is described by a smaller set of computational particles or "macro-particles"\cite{birdsall1991}. Such a model is acceptable as long as the statistical properties of the set of the macro-particles is close to that of the physical particles. It means that the moments resulting from integration in velocity space of the particle distribution and of the macro-particle distribution are very close, whatever their order. For the sake of readability, we shall use in the rest of the paper the word "particle" instead of "macro-particle". We remind that the moment of order $n$ of a distribution function is the integral over velocity space of this distribution function multiplied by the velocity at power $n$. Density, bulk velocity and pressure are the moments of order 0, 1 and 2 respectively, and are the most widely used.

A characteristic of the particles used in PIC codes is their finite size, whose spatial profile is given by a continuous function called the "assignment function" or "shape factor". This function has bounded support and it is thus zero outside of it. Importantly the size of the particle (their support) depends on the local spatial resolution as well as on the order of interpolation (see Ref. \cite{birdsall1991} for an extensive review). In the context of AMR codes this is crucial: when a particle moves from a position where the grid size is $\Delta$ to enter a refined region where the grid size is $\Delta / r$, the size of the particle should also decrease by a factor $r$, where $r$ is the refinement factor. When entering this region of smaller grid size, the "parent" particle has to be split into two or more "children" particles for at least two reasons: the first is to avoid spurious self-forces that are associated with the time derivative of the assignment function\sug{\cite{colella2010, vay2018}}, the second is to maintain a sufficiently large number of particles per cell\sug{\cite{winske2003, fujimoto2011, kolobov2016}, \cite{fujimoto2018}} in order to insure stability and accuracy. In that respect, particle splitting is also useful in codes using a uniform mesh. Therefore an important question that needs to be addressed is: how many children particles should the parent particle be split into? And as a corollary question, once this number is fixed, where should the children particles be dispatched, with which velocities and with which weights?
So far, only empirical answers to these questions can be found in the literature. Instead, this paper aims at addressing these questions rigorously by using an optimization technique that minimizes the difference between the assignment function of the parent particle and the sum of the assignment functions of the children particles.

The paper is structured as follows. In section 2, we review the important features of basis-splines (B-splines), as these are used to represent macro-particles in the vast majority of PIC codes. In section 3, we discuss the dichotomy between the exact solution of the splitting problem and the approximate one we are proposing. In section 4, we present some general considerations on splitting requirements, and the possible splitting patterns depending on the dimension of the problem. In section 5, we discuss the way to evaluate the accuracy of the method, and provide the optimal parameters of the split particles. In section 6, we discuss the consequences of the splitting on the level of density fluctuation on the refined grid. In section 7, we discuss possible implications of our conclusions for the inverse problem of merging process.

\section{B-splines for Particle-In-Cell method}

\begin{figure}[ht!]
\centering
\includegraphics[width=\textwidth]{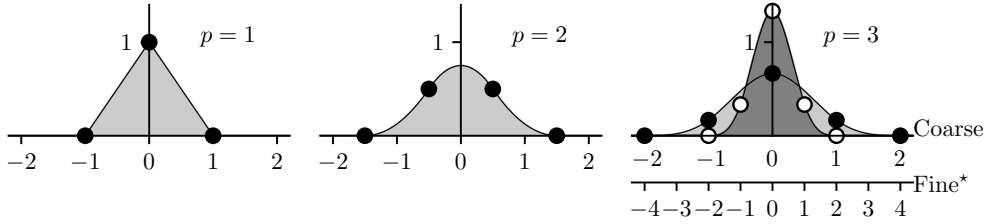}
\caption{Left panel depicts the B-spline for $p=1$, middle panel is the B-spline for $p=2$, and right panel is the B-spline for $p=3$. In each panels, light gray is used to represent the B-splines with a refinement factor $r=1$ (i.e. the parent particle). In addition, an example of a B-spline with a refinement factor $r=2$ (i.e. the child particle), is shown on right panel in dark gray. Black knots are for $r=1$ and white knots are for $r=2$.}\label{fig-splinedef}
\end{figure}

The particles' assignment functions used in PIC codes are generally B-spline functions \cite{lapenta2002}. These functions can be of any degree: the higher the degree, the larger the particle size, i.e. the length of the support of the assignment function \sug{and the smoother the particle shape}. Examples of B-splines of degree one to three are shown in Fig. \ref{fig-splinedef}. For clarity, we will treat in this paper the linear ($p=1$), quadratic ($p=2$) and cubic ($p=3$) cases. The $p=0$ case has been proven to be unstable \cite{birdsall1991}, while values of $p>3$ are very diffusive,  both cases are rarely used in actual simulations. A B-spline of degree $p$ is the union of $p+1$ polynomials of degree $p$ on $p+2$ knots. For the B-spline of degree $p=1$, the knots are $\{-1, 0, +1\}$, for $p=2$, they are $\{ -^3\!/\!_2, -^1\!/\!_2, +^1\!/\!_2, +^3\!/\!_2\}$ and for $p=3$, they are $\{-2, -1, 0, +1, +2\}$. The knots are shown in Fig. \ref{fig-splinedef} as black bullets on a grid of unitary mesh size.

To handle B-splines in a more general way, we shall use the notation $S^p(x)$ for a B-spline of degree $p$, defined on a grid of mesh size $\Delta$ centered at $x=0$ and $s^p(x-\delta)$ the B-spline of degree $p$ defined on a grid of mesh size $\Delta/r$  centered on $x=\delta$. These represent respectively the parent and children particles. \sug{By construction, B-splines have their integral equal to unity}. As an illustration, on the right panel of Fig. \ref{fig-splinedef}, the B-spline $s^3(x)$ for $r=2$ is depicted in dark gray, with its associated knots as white bullets. When a particle leaves a position where the grid size is $\Delta$ to reach a new one where the grid size is $\Delta/r$, the support of the B-splines has to be divided by $r$. \sug{For convenience, the coarse and fine grid have been displayed in the right panel of Fig. \ref{fig-splinedef},}

It is clear that when moving in a domain of smaller mesh size, the spatial profile of the assignment function of a particle can not be preserved if this particle is substituted by only one with a narrower support (see right panel of Fig. \ref{fig-splinedef}). The obvious consequence is that a parent particle needs to be split into several child particles. One may hope that the error introduced when splitting the parent particle somehow becomes unimportant when splitting many parent particles.
\sug{However, we shall demonstrate that this is \textit{not} the case and that the number and placement of child particles may have dramatic consequences on the conservation of the associated distribution function.}
For convenience we restate here the analytic form of B-splines with $\Delta$ = 1. The linear B-spline is given by

\begin{equation} \label{defl}
S^1(x) = \left| \begin{array}{ll}
1-|x| \hspace{4.0cm} & |x| \leq 1 \\
0 & |x| \geq 1
\end{array}
\right.
\end{equation}

\noindent
the quadratic B-spline is defined by

\begin{equation}\label{defq}
S^2(x) = \left| \begin{array}{ll}
{^3\!/\!_4}-x^2 \hspace{4.0cm} & |x| \leq {^1\!/\!_2} \\
{^1\!/\!_2}({^3\!/\!_2}-|x|)^2 & {^1\!/\!_2} \leq |x| \leq {^3\!/\!_2} \\
0 & |x| \geq {^3\!/\!_2}
\end{array}
\right.
\end{equation}

\noindent
and the cubic B-spline is defined by

\begin{equation}\label{defc}
S^3(x) = \left| \begin{array}{ll}
{^1\!/\!_2}|x|^3-x^2+{^2\!/\!_3} \hspace{8.0cm} & |x| \leq 1 \\
{^4\!/\!_3}(1-{^1\!/\!_2}|x|)^3 & 1 \leq |x| \leq 2 \\
0 & |x| \geq 2
\end{array}
\right.
\end{equation}

Without loss of generality, we shall focus in this paper on the special case (for the numerical values) where $\Delta=1$ for the parent and $r=2$ for the children, but we shall keep $\Delta$ and $r$ in the notations. A particle $k$ is thus defined by its weight $w_k$, position $\mathbf x_k$ and velocity $\mathbf v_k$, so the continuous density function at position $\mathbf x$

\begin{equation}
n(\mathbf x) = \int_{\mathbb R^3} f(\mathbf x, \mathbf v, t) \: d \mathbf v
\end{equation}

\noindent
where $f(\mathbf x, \mathbf v, t)$ is the distribution function, is approximated by the sum

\begin{equation}
n(\mathbf x) \sim \sum_{k=1}^M w_k S^p (\mathbf x - \mathbf x_k)
\end{equation}

In the 2D and 3D cases, the assignment function is the product of the 1D assignment function defined for each directions, for example:

\begin{equation}
S^p(\mathbf x - \mathbf x_k) = S^p(x - x_k) S^p(y - y_k) S^p(z - z_k) \label{eq-spk}
\end{equation}

\section{Exact and approximate solution}\label{sec-solution}

In a pioneering study by Lapenta \cite{lapenta2002}, it was emphasized that the rezoning of particles (splitting or merging) can not be exact for \textit{uniform} meshes. While this assertion is true for a uniform grid, it no longer applies if the children are dispatched to a mesh of different grid size from that of their parent. In this case the size of the support of the assignment function of the children is smaller (by a factor $r$) than that of their parent. Indeed, an exact solution exists when $r$ is an integer\cite{bornemann2013}. However when applied to PIC codes, the solution comes at the cost of a large number of split particles: given the dimension $d$ of the grid and a refinement factor $r=2$, the number of children $N$ from a single parent is $(p+2)^d$. In three-dimension $d=3$, one gets 64 children from a single parents, which rapidly becomes prohibitive in practical applications.

\begin{figure}[ht!]
\centering
\includegraphics[width=\textwidth]{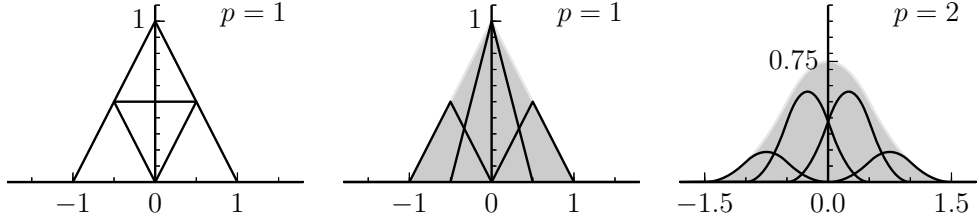}
\caption{Left panel depicts for $p=1$ the parent and the 4 children for the exact solution. Middle panel is the same as left panel, where the two children at the origin can be viewed as a single child of weight $ {^1\!/\!_2} $. Right panel has the same format as middle panel for $p=2$. Each panel is for $d=1$.}\label{fig-bornemann}
\end{figure}

For completeness we now discuss the exact splitting for a 1D case and indicate how to extend it to higher dimensions. Examples of exact splitting in 1D are illustrated in Fig. \ref{fig-bornemann}. On the left panel, the B-spline of the parent particle, as well as the B-splines of the four children are of the same degree, namely $p=1$. This example illustrates that an exact solution exists with four children of the same weight $ {^1\!/\!_4} $ (the weight of the parent particle being 1). The two children located at the origin are equivalent to a single child of weight $ {^1\!/\!_2} $, so the exact solution needs in principle $N=3$ children for $p=1$. This is depicted in the middle panel of Fig. \ref{fig-bornemann}. For $p=2$, the exact solution requires $N=4$ children and it is illustrated in the right panel of Fig. \ref{fig-bornemann}.

In section 5, we provide the parameters for the exact splitting in one dimension, for B-splines of degree $p$ = 1, $p$ = 2 and $p$ = 3. More specifically, we provide the $w_i$ and $\delta_i$ satisfying the relation

\begin{equation}
S^p(x) = \sum_{i=1}^N w_i s^p(x-\delta_i) \label{eq-Spsp}
\end{equation}



The 2D and 3D generalization is straightforward using Eq. (\ref{eq-spk}).

For the approximate splitting, the problem comes down to fix $N$, the number of split particles, as well as their weights $w_i$ and positions $\delta_i$, in order to get an assignment function of the children as close as possible to the one of the parent at every points of their support.

A solution to this problem is to split the parent into $N$ children, all of them having the same velocity as the parent.
In this case, all the moments of the distribution function (density, velocity, pressure, heat flux, $\dots$) associated with the collection of children can be made the same as the moments of the distribution of the parents, provided it is the case for the density profile (the first moment). Then, for a given value of the refinement factor $r$, and a given value of the number of children $N$, the problem reduces to finding the position and an associated weight for each child that minimizes the difference between the moments of the distributions calculated on the coarse and fine grids.

To our knowledge, in AMR codes using macro-particles (e.g. \cite{fujimoto2006}, \cite{muller2011}, \cite{innocenti2013}) the splitting of particles is neither exact nor optimized. In addition, these methods are all based on split particles having the same velocity as their parent. While arbitrary, this choice simplifies the problem and it is the one we adopt here. We now discuss in more details the splitting methods existing in the literature.
\begin{itemize}
    \item In the study of Fujimoto\cite{fujimoto2006}, the symmetry in the 2D case is preserved with four children. The children are located at a distance $\delta^{\star} = \Delta/\sqrt{N_{PPC}}$ from the parent where $\Delta$ is the grid size associated to the parent and $N_{PPC}$ is the number of particle in the cell where the parent is located. With $\Delta = 1$ and $N_{PPC} \sim 100$ (which is a commonly used value), one obtains $\delta^{\star} \sim 0.1$.
    \item In the study by Innocenti et al.\cite{innocenti2013} the symmetry is also preserved but the number of split particles depends on the dimension $d$. Eq. (39) of Ref. \cite{innocenti2013} shows that, in our notation, these children are uniformly dispatched along each direction with a spacing $ {^1\!/\!_2} (\Delta/r) = {^1\!/\!_4}$ for $\Delta = 1$ and $r=2$. Hence, in the 2D case, $\delta^{\star} = 0.25$ for four children (keeping in mind that a larger, even number of split particles can be used with this same $\delta^{\star}$ value).
    \item In the study of Muller et al.\cite{muller2011} the symmetry is not preserved. The parent is split into two children independently of the dimension of the problem. Furthermore, the position $\delta^{\star}$ of the two children is not precisely given, except that the independent shift of each of them, relative to the parent, is ``small''. \sug{We shall consider that $\delta^{\star}$ is of the order of 0.1}. 
\end{itemize}

\sug{The star notation for $\delta^{\star}$ is intended to outline the fact that these values are defined on the fine grid (see right panel of Fig. \ref{fig-splinedef}) and not on the coarse one}.

\sug{Making a clear difference with the choices of these studies, we stress that it is crucial to ensure that the assignment function of the parent and of the set of children are as close as possible, at every points of their support. Hence, the optimization process must rely on a procedure considering these assignment functions as continuous functions, and not solelly considering a finite number of their moments (density, momentum, energy) as in the studies cited above.}


\begin{figure}[ht!]
\centering
\includegraphics[width=\textwidth]{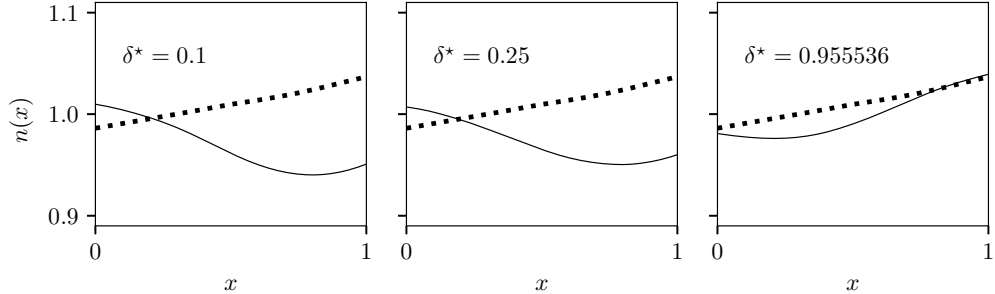}
    \caption{Density profiles $n(x)$ in a 2D cell of width $\Delta=1$ for a set of 100 parents (thick dotted lines), and a set of 400 children deposited at $\pm \delta^{\star}$ from their respective parent in each direction. From left to right panels, the $\delta^{\star}$ values are 0.1, 0.25 and 0.955536.}\label{fig-flu3}
\end{figure}

To make clear the close relation between the position of the children relative to their parent and the associated changes in the density profile, we randomly distribute 100 parents in a single, 2D cell of unitary mesh size (some more particles are also deposited on the adjacent cells to avoid the drop of density on the border of the cell). \sug{We Focused on the 2D case as this case is nowadays more prevalent than the 1D case. Nonetheless, to ease the representation, we decided to display a line out along the $X$ direction, obtained at a given $Y$ position}. As an illustration, we take an assignment function of degree $p=1$. The density profile $n(x)$, for a given $y$,  calculated for this set of parents is depicted in Fig. \ref{fig-flu3} using a thick dotted line. The fluctuations in the density (i.e. the fact that the density is not equal to \sug{1.0} in the whole cell) are the result of the random distribution of these parents in each cell. As an illustrative example, we represent the splitting process needed for a refinement factor $r=2$ and dimension $d=2$, using $N = 4$ children. In order to conserve mass, each parent is then split into four children of equal weight, $w_i$ = $^1\!/\!_4$. As for the parents, the children assignment function is of degree $p=1$, however because of the refinement, their support is two times smaller than that of the parents. The position of the children is shifted with respect to the position of their parent by \sug{$\pm \delta^{\star}$} in each direction.
The density profile of this set of children, resulting from the superposition of the 400 children, are depicted in solid black lines in the three panels of Fig. \ref{fig-flu3}, for \sug{$\delta^{\star} = 0.1$ (left panel), $\delta^{\star} = 0.25$ (middle panel) and $\delta^{\star} = 0.955536$ (right panel).}

The first two panels correspond to the two values of \sug{$\delta^{\star}$} used in previous studies\cite{fujimoto2006}, \cite{muller2011}, \cite{innocenti2013}. The last panel shows the method presented in this paper. It is clear from Fig. \ref{fig-flu3} that the choice of \sug{$\delta^{\star}$} is crucial
\sug{in order for the density distribution of the children to be as close as possible to that of the parents. In addition, the density profiles show larger total variation depending on the choice of $\delta^{\star}$, which may have consequences on the density level of fluctuations. This point will be deepened in section \ref{sec-accuracy}.}

\section{Constraints on the splitting strategy}

The problem treated in this paper is to determine for a single parent of weight unity located at the origin, the number of children $N$, their weight $w_i$ and location $\delta_i^{\star}$ for $i = \{1, 2, \dots N\}$ so that the associated assignment functions are as close as possible. If the dimension $d$ of the problem is larger than one, then $\delta_i^{\star}$ are vectors. It is reasonable to think that the larger $N$, the smaller the associated error. The yet unspecified definition of this error will be discussed in section \ref{sec-accuracy}, but it essentially quantifies how far the assignment function of the children are from the assignment function of the parent. In order to control the accuracy we want to reach, we keep $N$ as a free parameter. We focus on the position of the children because, as already mentioned, their velocity is the one of their parent.

\begin{figure}[ht!]
\centering
\includegraphics[width=\textwidth]{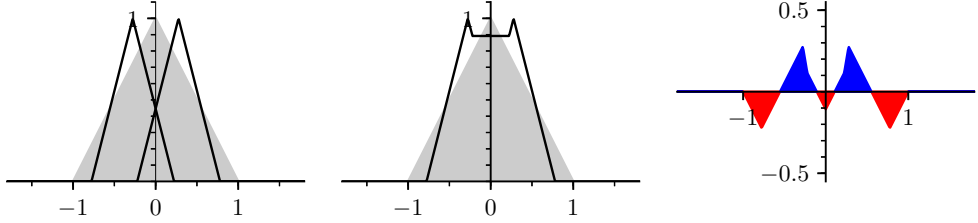}
\caption{Left panel depicts for $p=1$ the parent and the two children located at $\delta = \pm 0.277$ (normalized to the coarse grid mesh $\Delta$). Middle panel is the same as left panel where only the sum of the two children is displayed. Right panel is the difference between the parent and the sum of the two children (blue when the sum is positive and red when it is negative).}\label{fig-diff}
\end{figure}

\sug{As already said, $\delta$ and $\delta^{\star}$ are the position of the children, relative to their parent, defined on the coarse (of grid size $\Delta$) and fine (of grid step $\Delta/r$) grid, respectively. Hence, $\delta^{\star} = r \delta$.}
\sug{Yet, we also use $\delta$, defined on the coarse grid, in some cases, as in Eq. \ref{eq-Spsp}.}
From a technical point of view, refined particles are dispatched on the refined grid, so their positions have to be defined on this refined grid. The method for $r\neq 2$ is also applicable, but needs to recalculate the $w_i$ and $\delta_i^{\star}$ as discussed in the next section. Such calculations can be done using the dedicated optimization code \cite{splitpic}.

Among all the sets of $w_i$ and $\delta_i^{\star}$, we want to pick the ones for which the assignment function of the children is the closest to the one of the parent. This is illustrated in Fig. \ref{fig-diff} for the simple 1D case, with $p=1$ and $N=2$, where $N$ is the number of split particles on the fine grid for a single particle on the coarse grid. On the left panel, the two children are depicted by a black line, while the parent is in gray. One notes that the support of the assignment function of the children being half the one of the parent, its maximum value is twice the one of the parent because B-splines are normalized functions. In the left panel of Fig. \ref{fig-diff}, the two children are deposited with a weight $^1\!/\!_2$ in order to conserve mass. On the middle panel, we display the sum of the assignment functions of the two children (solid black line), as well as the one of the parent (also in gray). On the right panel, we display the difference between the assignment function of the parent and the one of the children. Red is used when this difference is positive and blue is used when it is negative. The best solution is the one for which the total (red plus blue) surface is as small as possible. We can then define the cost function in order to determine, for each $d$, $p$ and $N$ values, the $w_i$ and $\delta_i$ for each child $i$. For the 1D case, the cost function is:

\begin{equation} \label{eq-qdef}
Q_N^p = \int_{\mathbb R} \left| S^p(x) - \sum_{i=1}^{N} w_i s^p(x-\delta_i) \right|^2 dx
\end{equation}

For the sake of readability, we define the assignment function of the set of children as the sum of the assignment functions of each of them. The optimization problem we need to solve is to find the best values of the free parameters ($w_i$ and $\delta_i$) for a given number of children $N$, to minimize the cost function defined as the difference between the assignment functions of the parent and the one of the children. This is an optimization problem as assignment functions are continuous functions and the number of free parameter is finite.

We emphasize that the evenness/oddness of the number of split particles, $N$, is constrained by symmetry considerations. B-splines are by construction even functions. For a parent located at the origin, one can focus on the positive half of the domain. Let's consider we know the weights and their locations (eventually in the half negative domain) for the set of children that minimizes the cost function. Then, by evenness of B-splines, this same set will also minimize this difference in the negative half domain, provided the sign of the $\delta_i$ are changed accordingly. As a consequence, the best solution will be reached for an even number of children. In fact, adding a single child at the origin (collocated with the parent) won't modify the evenness requirement discussed above.

\begin{figure}[ht!]
\centering
\includegraphics[width=\textwidth]{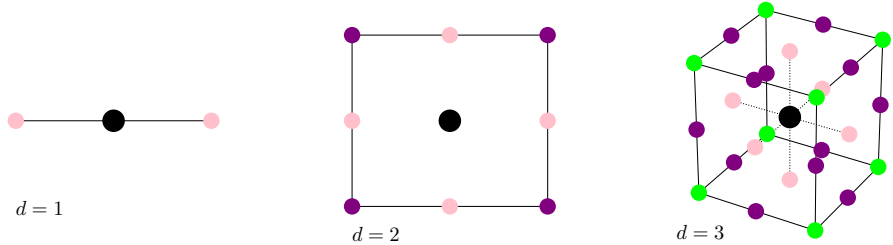}
    \caption{The parent is depicted by the large black bullet. Depending on the pattern \sug{type,} the children are depicted in pink, purple and lime. Left panel is the $d=1$ case, with two children. Middle panel is the $d=2$ case, with four children in two \sug{types of} patterns. Right panel is the $d=3$ case, with six, eight or twelve children in three \sug{types of} patterns.}\label{fig-where}
\end{figure}

Given the evenness considerations of assignment function just discussed, we now need to determine the spatial pattern to be followed to dispatch particles. These patterns do not depend on the degree $p$ of the B-splines, but importantly, they depend on the dimension $d$ of the problem. For the 1D case, the solution is quite simple: the two children have the same weight and are located at $\pm \delta$ around the parent \sug{in order to preserve the evenness of the assignment function}. This pattern is depicted in the left panel of Fig. \ref{fig-where}, where the parent is represented by a black bullet and the two children by pink bullets. For a larger (even) number of children, all couples of children are dispatched in a similar way. We shall call this, pattern \sug{of type} 1. A single child can also be dispatched at the origin (black position in Fig. \ref{fig-where}); this is pattern \sug{of type} 0. The \sug{number associated to the type of} these pattern will become clear at the beginning of next section.

For the 2D case, in order to preserve spatial symmetry (which is a necessary condition to reach the best solution), there are two ways to split a single parent in four children of equal weight : (i) the four particles are located at ($x=\pm \delta$, $y=\pm \delta$) or (ii) two are located at ($x=0$, $y=\pm \delta$) and the two others are located at ($x=\pm \delta$, $y = 0$). This first solution is depicted by purple bullets in middle panel of Fig. \ref{fig-where} and is called pattern \sug{of type} 2, while the second one is depicted by pink bullets and is called pattern \sug{of type} 1. One can notice that pink children are located at a distance $\delta$ from its parent while the purple ones are located at $\delta\sqrt2$. This farther location will have consequences on the $\delta$ value as well as on the accuracy that we will discuss in next section. If a single child of weight $w_0$ (associated to pattern \sug{of type} 0 as in the 1D case) is added at the origin, then the weight $w_1$ of the four other children is such as $w_0 + 4 w_1 = 1$. The $N=4$ case has then one free parameter (the $\delta$ value), while the $N=5$ case has two (the $\delta$ value and one of the two weights).

The same symmetry considerations apply to the 3D case. A single parent can be split in six, eight or twelve children: (i) the six particles are located at ($x = \pm \delta$, $y = 0$, $z = 0$) and the two associated circular permutation of directions, depicted by pink bullets in right panel of Fig. \ref{fig-where} (pattern 1), (ii) the eight particles are located at ($x=\pm \delta$, $y=\pm \delta$, $z = \pm \delta$), depicted by \sug{lime} bullets in Fig. \ref{fig-where} (pattern 3) or (iii) the twelve particles are located at ($x=\pm \delta$, $y=\pm \delta$, $z = 0$) and the two associated circular permutation of the directions, depicted by \sug{purple} bullets in Fig. \ref{fig-where} (pattern 2). As depicted in the right panel of Fig. \ref{fig-where}, adding a single child at the origin (pattern 0 as in the 1D and 2D cases) also illustrates the $N=7$, $N=9$ and $N=13$ cases. We should also mention that there are two different cases associated to $N=12$ (or $N=13$) : twelve purple children (one set of pattern 2) or two sets (of pattern 1) of six pink children (plus a single child at origin for $N=13$).

\section{Accuracy of the splitting strategy}\label{sec-accuracy}

\sug{For a subset of children, the associated pattern (defining the number and space distribution of children) is called $\tau_j$ where $j$ is the index of this subset}.
Naming $N_j$ the number of children \sug{for the $j$ subset of children with pattern $\tau_j$}, the assignment function of this subset of particles is noted

\begin{equation} \label{eq-stilda}
T_j^p(\mathbf r) = \sum_{i=1}^{N_j} s^p(\mathbf r- \boldsymbol{\delta}_i)
\end{equation}

\sug{In order to make the notations as explicit as possible, index $j$ of $T$ is the index of the subset of children (associated to type $\tau_j$), while index $i$ refers to a given child of this subset}.
The meaning of \sug{the type $\tau_j$} index can be clearly explained for the 3D case. A child \sug{$i$ belonging to the pattern of type $\tau_j$} is dispatched at a position

\begin{equation}\label{eq-deltai}
\boldsymbol{\delta}_i = a \delta_j \hat{\mathbf x} + b \delta_j \hat{\mathbf y} + c \delta_j \hat{\mathbf z}
\end{equation}

\noindent
where $a$, $b$ and $c$ belong to \{-1, 0, +1\} and $\hat{\mathbf x}$, $\hat{\mathbf y}$, $\hat{\mathbf z}$ are unit vectors. \sug{It is so because the $\boldsymbol{\delta}_i$ differ one from the other because of the different \{$a$, $b$, $c$\} values, but rely on the same $\delta_j$ value for the pattern of type $\tau_j$.}
\sug{The type of the pattern $j$ is defined as $\tau_j = a^2+b^2+c^2$}
so the associated pattern contains all possible values of \{$a$, $b$, $c$\} satisfying this relation. In Fig. \ref{fig-where}, pink, purple and lime pattern are then associated to \sug{$\tau_j$} = 1, 2 and 3, respectively. The naming of the $\tau_j=0$ pattern is then obvious.
\sug{The full knowledge of a subset $j$ of children is then given by the triplet ($\tau_j$, $w_j$, $\delta_j$).}
This can be extended straightforwardly to 2D and 3D cases.

\noindent
Eq. (\ref{eq-qdef}) can hence be written in a more compact way as

\begin{equation} \label{eq-qdefbis}
    Q_N^p = \int_{\mathbb R^d} \left| S^p(\mathbf r) - \sum_{j=1}^{M} w_j T_j^p(\mathbf r) \right|^2 d \mathbf r
\end{equation}

\noindent
In Eq. (\ref{eq-qdefbis}), \sug{$M$ is the number of subsets (each one being associated to a pattern type $\tau_j$) so that the total number of childern in Eq. (\ref{eq-qdef}) is given by $N = \sum_{j=1}^{M} N_j$ }.
\sug{It is clear that a pattern type $\tau_j$ for $j \in[1, M]$ can have zero or multiple occurences, with different $w_j$ and $\delta_j$ values (see right panel of Fig. \ref{fig-bornemann}).}
Because of the piecewise definition of B-splines, the analytical form of this cost function is not easy to obtain. However, optimal parameters can be obtained numerically and tabulated so to be used in codes at run time.

\begin{figure}[ht!]
\centering
\includegraphics[width=\textwidth]{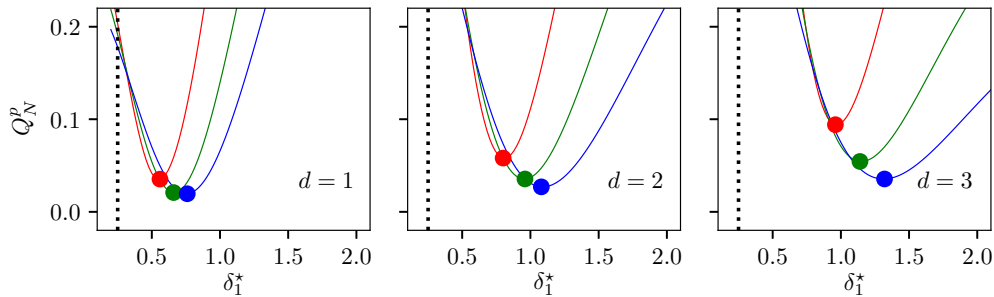}
\caption{$Q_N^p$ value given by Eq. (\ref{eq-qdef}) depending on the $\delta_1$ value. This correspond to a single pattern $j=1$, indicated by pink positions in Fig. \ref{fig-where}. Solid red, green and blue are for $p=1, 2, 3$ respectively. Vertical dotted lines at 0.25 is the solution proposed by Innocenti et al.\cite{innocenti2013}, which is off the chart for $d=3$.}\label{fig-qDelta}
\end{figure}

We calculate $Q_N^p$ defined by Eq. (\ref{eq-qdef}) for each combinations of $p=1, 2, 3$ and $d=1,2,3$. The values obtained for the single free parameter \sug{$\delta_1^{\star}$ (reminding that the star notation is associated to the fact that this value is defined on the refined grid)} are depicted in Fig. \ref{fig-qDelta}. The left panel is for $d=1$ and two children \sug{($N_1 = 2$)}, the middle panel is for $d=2$ and four children \sug{($N_1 = 4$)}), and the right panel is for $d=3$ and six children \sug{($N_1 = 6$)}. The curves correspond to the \sug{$\tau_1=1$} pattern, indicated by the pink positions in Fig. \ref{fig-where}, whatever the value of $d$. In each panels, the red curve corresponds to $p=1$, green is for $p=2$ and blue is for $p=3$. One notes that it exists for each set of $p$ and $d$ values, an absolute minimum and no local minima, i.e. an optimal \sug{$\delta_1^{\star}$} value to dispatch the split particles. In Fig. \ref{fig-qDelta}, we also displayed, using bullets, the positions of the minima for each curves.

The minimum values $Q_N^p$ are increasing with $d$, but decreasing with $p$. This last point results from the fact that the larger $p$, the smaller the total variation of the function. Moreover, the optimum $\delta_1^{\star}$ value is increasing with $d$, whatever the $p$ value. This results from the increasing values of \sug{$N_1$} with $d$, hence the decrease of the weight of each child: being lighter, they need to be dispatched further from the parent to fulfill the tail of the assignment function  of the parent. Furthermore, the support of a B-spline is increasing with $p$. The support of a child being half the one of its parent, this child has to be dispatched farther from its parent. As a consequence, $\delta_1^{\star}$ increases with $p$.

As already discussed in section \ref{sec-solution}, the suggested $\delta_1^{\star}$ value in Ref. \cite{fujimoto2006} and  \cite{muller2011} are close to 0.1. This value is off chart, very far from the optimal $\delta_1^{\star}$ value and it is associated to a very large error. In Ref. \cite{innocenti2013}, the suggested value of $\delta_1^{\star}$ is 0.25 (independently of the $p$ and $d$ values), which is displayed in left and middle panels of Fig. \ref{fig-qDelta} with dotted vertical lines (this value being out of the scaling for the right panel). It clearly appears that these two values are not the best one in order to satisfy our accuracy criterion. Moreover, as can be observed on Fig. \ref{fig-flu3}, these $\delta_1^{\star}$ values also increase the level of fluctuations of the density (and higher order moments) profile, which generally compromises the stability of the code. The numerical increase of the level of fluctuation will be provided and discussed in the next section.

With the power 2 in Eq. (\ref{eq-qdef}), one recognizes the use of the  $L^2$ norm. \sug{We tried to use a different power, namely 1 and 3. The derivative of $Q_N^p$ is smaller for higher power values in the norm, meaning that the associated minimum value is approximately at the same position, but in a shallower potential well. The good stability for the location of the minima then suggests that the results given here do not depend on the choice of the norm.}

\begin{figure}[ht!]
\centering
\includegraphics[width=\textwidth]{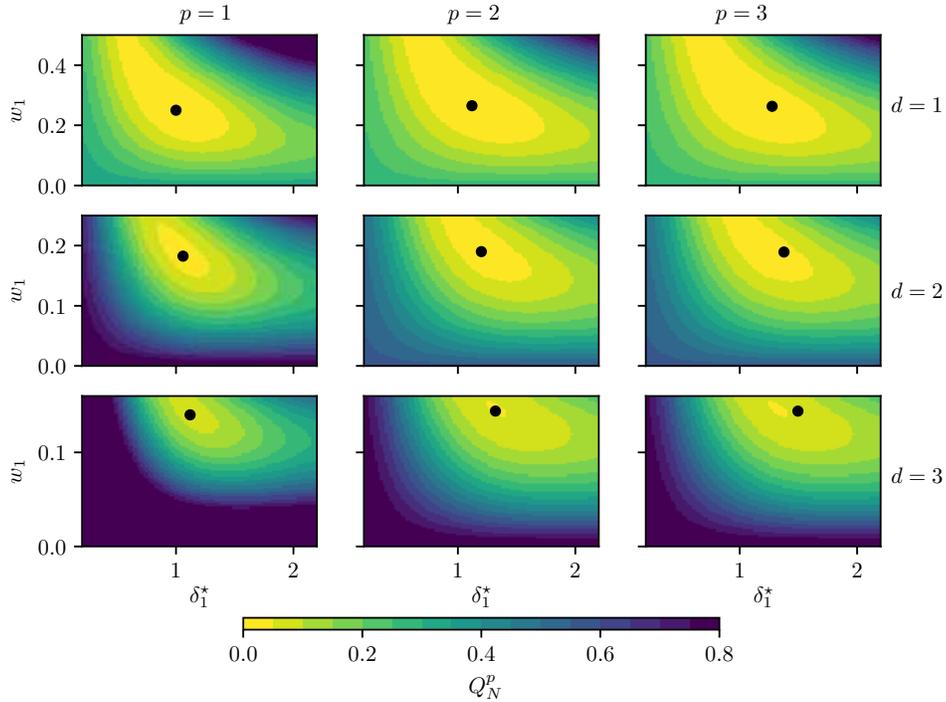}
\caption{$Q_N^p$ value given by Eq. (\ref{eq-qdef}) depending on the $w_1$ and $\delta_1$ values. This situation correponds to two pattern $j=0$ and $j=1$. Rows are for $d=1, 2, 3$ from top to bottom, respectively, and columns are for $p=1, 2, 3$ from left to right, respectively. We use a different scaling for the $Y$-axis, depending on the dimension $d$, but the same at a given $d$ value whatever the $p$ value. The reason is that the relation $w_0 = 1-N w_1$ has to be satisfied as well as the positiveness of the $w_i$. As the number of children $N$ increases with $d$, the range of possible values for $w_1$ thus decreases. The scaling of the $X$-axis is the same for all the nine panels, but the origin of the $X$-axis depends on the degree $p$ of the B-splines: the larger $p$, the larger the origin of the $X$-axis. As for Fig. \ref{fig-qDelta}, this is a consequence of the increasing support of B-splines with their degree $p$.}\label{fig-qWDelta}
\end{figure}

We now carry out the same optimization with two degrees of freedom, $w_1$ and $\delta_1^{\star}$. In this case, \sug{$N_1$} children (with a pattern $\tau_1$ = 1) of weight $w_1$ are dispatched according to the pink positions in Fig. \ref{fig-where} with the associated $\delta_1^{\star}$ parameter, and a single child (for the $\tau_2$ = 0 pattern) of weight $w_2 = 1 - (N_1-1) w_1$ is located at the origin. Results are displayed in Fig. \ref{fig-qWDelta}.  For clarity, we use the same color code for each panel, ranging from 0 (\sug{yellow}) to 0.8 (\sug{dark blue}), \sug{so $Q_N^p$ values larger than 0.8 are saturated in dark blue}. It is clear that, as for Fig. \ref{fig-qDelta}, the minimum value of $Q_N^p$ increases with $d$, decreases with $p$, and is \sug{narrower} as $d$ increases. \sug{Depending on the dimension $d$ and order $p$,  the potential wells, for each cases, have a different location and depth. Furthermore, we adapt for each panels the range of the $w_1$ values to focus on the region preserving the positivity of $w_2$. The most important feature from these panels is that with two degrees of freedom, the property of having a single absolute minimum is preserved, whatever the $p$ and $d$ values. We can conjecture it is so for larger \sug{$M$} value, even if this point is not mandatory for the numerical resolution of this problem.}

We pursue these calculations to find the $w_j$'s and $\delta_j^{\star}$'s values that minimize $Q_N^p$ for larger values of $N$ using Eq. (\ref{eq-qdefbis}). We report these $w_j$ values in Tab. \ref{tab1dw}, \ref{tab2dw} and \ref{tab3dw} for $d$ = 1, $d$ = 2 and $d$ = 3, respectively. The associated $\delta_j^{\star}$ are in Tab. \ref{tab1dd}, \ref{tab2dd} and \ref{tab3dd} for $d$ = 1, $d$ = 2 and $d$ = 3, respectively. Note that in Tab. \ref{tab1dw} to \ref{tab3dd}, the first row contains 
\sug{the total number $N$ of children for this configuration, while the second contains the type $\tau_j$ of the pattern used for this subset $j$ (one per line) of children that have to be considered in the summation of Eq. (\ref{eq-qdefbis}).}
For the interested reader, the associated python code for this optimization\cite{splitpic} (using the NLopt optimization library\cite{nlopt}) can be easily used and extended for larger $p$, different $r$ and/or larger $N$. It is also very important (for the implementation on an AMR code) to remind that the $\delta_j^{\star}$ values are given on the refined mesh in units of $\Delta/r = {^1\!/\!_2}$.

\begin{table}
\centering
\begin{tabular}{l c c c c}
\hline
\hline
N & $\tau$ & \hspace{0.0cm} $p=1$ & \hspace{0.0cm} $p=2$ & \hspace{0.0cm} $p=3$ \\
\hline
\hline
2 & 1  &     0.5   & 0.5        & 0.5 \\
\hline
3 & 0  & \bf{0.5}  & 0.468137   & 0.473943 \\
  & 1  & \bf{0.25} & 0.265931   & 0.263028 \\
\hline
4 & 1 &            & \bf{0.375} & 0.364766 \\
  & 1 &            & \bf{0.125} & 0.135234 \\
\hline
5 & 0 &            &            & \bf{0.375} \\
  & 1 &            &            & \bf{0.250} \\
  & 1 &            &            & \bf{0.0625}\\
\hline
\hline
\end{tabular}
\caption{\label{tab1dw} Values of $w_i$(s) for $p$ = 1, 2 and 3 and $d$ = 1. Bold values are for exact splitting.}
\end{table}

\begin{table}
\centering
\begin{tabular}{l c c c c}
\hline
\hline
N & $\tau$ & \hspace{0.0cm} $p=1$ & \hspace{0.0cm} $p=2$ & \hspace{0.0cm} $p=3$ \\
\hline
\hline
2 & 1  & 0.551569   & 0.663959   &  0.752399\\
\hline
3 & 0  & ---        & ---        & --- \\
  &  1 & \bf{1.0}   & 1.112033   & 1.275922 \\
\hline
4 & 1 &             & \bf{0.5}   & 0.542949 \\
  & 1 &             & \bf{1.5}   & 1.664886 \\
\hline
5 & 0 &             &            & --- \\
  & 1 &             &            & \bf{1.0}   \\
  & 1 &             &            & \bf{2.0}   \\
\hline
\hline
\end{tabular}
    \caption{\label{tab1dd} Values of $\delta_i^{\star}$(s) for $p$ = 1, 2 and 3 and $d$ = 1, \sug{defined on the refined grid}. Bold values are for exact splitting.}
\end{table}

\begin{table}
\centering
\begin{tabular}{l c c c c}
\hline
\hline
N & $\tau$ & \hspace{0.0cm} $p=1$ & \hspace{0.0cm} $p=2$ & \hspace{0.0cm} $p=3$ \\
\hline
\hline
    4 & 1  & 0.25$^{\dagger}$      & 0.25$^{\dagger}$     & 0.25$^{\dagger}$ \\
\hline
4 & 2  & 0.25      & 0.25     & 0.25 \\
\hline
    5 & 0  & 0.270426$^{\dagger}$  & 0.239166 & 0.242666 \\
    & 1  & 0.182394$^{\dagger}$  & 0.190209 & 0.189333 \\
\hline
    5 & 0  & 0.239863  & 0.210694$^{\dagger}$ & 0.22294$^{\dagger}$ \\
    & 2  & 0.190034  & 0.197327$^{\dagger}$ & 0.194265$^{\dagger}$ \\
\hline
8 & 1  & 0.179488  & 0.178624 & 0.179318 \\
  & 2  & 0.070512  & 0.071376 & 0.070682 \\
\hline
9 & 0 & \bf{0.25}  & 0.213636 & 0.218605 \\
  & 1 & \bf{0.125} & 0.126689 & 0.126871 \\
  & 2 & \bf{0.0625}& 0.069902 & 0.068477 \\
\hline
\hline
\end{tabular}
    \caption{\label{tab2dw} Values of $w_i$(s) for $p$ = 1, 2 and 3 and $d$ = 2. Bold values are for exact splitting. \sug{The dagger exponent indicate the worst pattern between the two possible cases for $N$ = 4 and $N$ = 5.}}
\end{table}

\begin{table}
\centering
\begin{tabular}{l c c c c}
\hline
\hline
N & $\tau$ & \hspace{0.0cm} $p=1$ & \hspace{0.0cm} $p=2$ & \hspace{0.0cm} $p=3$ \\
\hline
\hline
    4 & 1  & 0.807166$^{\dagger}$  & 0.955536$^{\dagger}$ & 1.089404$^{\dagger}$ \\
\hline
4 & 2  & 0.571783  & 0.683734 & 0.776459 \\
\hline
    5 & 0 & ---$^{\dagger}$        & ---      & --- \\
    & 1 & 1.053876$^{\dagger}$   & 1.203227 & 1.376953 \\
\hline
    5 & 0 & ---        & ---$^{\dagger}$      & ---$^{\dagger}$ \\
    & 2  & 0.721835  & 0.83043$^{\dagger}$  & 0.956756$^{\dagger}$ \\
\hline
8 & 1  & 0.700909  & 0.828428 & 0.942365 \\
  & 2  & 1.05786   & 1.236701 & 1.423324 \\
\hline
9 & 0 & ---        & ---      & --- \\
  & 1 & \bf{1.000} & 1.105332 & 1.267689 \\
  & 2 & \bf{1.000} & 1.143884 & 1.315944 \\
\hline
\hline
\end{tabular}
    \caption{\label{tab2dd} Values of $\delta_i^{\star}$(s) for $p$ = 1, 2 and 3 and $d$ = 2, \sug{defined on the refined grid}. Bold values are for exact splitting. \sug{The dagger exponent indicate the worst pattern between the two possible cases for $N$ = 4 and $N$ = 5.}}
\end{table}

\begin{table}
\centering
\begin{tabular}{l c c c c}
\hline
\hline
N & $\tau$ & \hspace{0.0cm} $p=1$ & \hspace{0.0cm} $p=2$ & \hspace{0.0cm} $p=3$ \\
\hline
\hline
6  & 1 & 0.166666 & 0.166666 & 0.166666 \\
\hline
7  & 0 & 0.155626 & 0.13594  & 0.136213 \\
   & 1 & 0.140729 & 0.14401  & 0.143964 \\
\hline
8  & 3 & 0.125    & 0.125    & 0.125    \\
\hline
9  & 0 & 0.129097 & 0.119495 & 0.128032 \\
   & 3 & 0.108863 & 0.110063 & 0.108996 \\
\hline
12 & 2 & 0.083333 & 0.083333 & 0.083333 \\
\hline
13 & 0 & 0.1552   & 0.137335 & 0.142364 \\
   & 2 & 0.0704   & 0.071889 & 0.07147  \\
\hline
14 & 1 & 0.101754 & 0.094953 & 0.096661  \\
   & 3 & 0.048684 & 0.053786 & 0.052504  \\
\hline
15 & 0 & 0.137684 & 0.143053 & 0.143017  \\
   & 1 & 0.076854 & 0.056257 & 0.062838  \\
   & 3 & 0.050149 & 0.064926 & 0.059994  \\
\hline
18 & 1 & 0.082439 & 0.077118 & 0.078179  \\
   & 2 & 0.042114 & 0.044775 & 0.044244  \\
\hline
19 & 0 & 0.128816 & 0.090629 & 0.093679  \\
   & 1 & 0.062366 & 0.061106 & 0.061855  \\
   & 2 & 0.041416 & 0.045228 & 0.044599  \\
\hline
20 & 2 & 0.064204 & 0.065395 & 0.065154  \\
   & 3 & 0.028694 & 0.026908 & 0.02727   \\
\hline
21 & 0 & 0.135727 & 0.110848 & 0.116674  \\
   & 2 & 0.061347 & 0.062333 & 0.061983  \\
   & 3 & 0.016014 & 0.017645 & 0.017441  \\
\hline
26 & 1 & 0.082117 & 0.078837 & 0.079616  \\
   & 2 & 0.031737 & 0.03104  & 0.031146  \\
   & 3 & 0.015806 & 0.019311 & 0.018569  \\
\hline
27 & 0 &\bf{0.125}& 0.099995 & 0.104047  \\
   & 1 &\bf{0.0625}& 0.055301 & 0.05564  \\
   & 2 &\bf{0.03125}& 0.035568 & 0.035385 \\
   & 3 &\bf{0.015625}& 0.017672 & 0.017187 \\
\hline
\hline
\end{tabular}
\caption{\label{tab3dw} Values of $w_i$(s) for $p$ = 1, 2 and 3 and $d$ = 3. Bold values are for exact splitting.}
\end{table}

\begin{table}
\centering
\begin{tabular}{l c c c c}
\hline
\hline
N & $\tau$ & \hspace{0.0cm} $p=1$ & \hspace{0.0cm} $p=2$ & \hspace{0.0cm} $p=3$ \\
\hline
\hline
6  & 1 & 0.966431 & 1.149658 & 1.312622 \\
\hline
7  & 0 & ---      & ---      & --- \\
   & 1 & 1.121649 & 1.310004 & 1.495565 \\
\hline
8  & 3 & 0.584015 & 0.700806 & 0.79718 \\
\hline
9  & 0 & ---      & ---      & --- \\
   & 3 & 0.664932 & 0.785409 & 0.901924 \\
\hline
12 & 2 & 0.74823  & 0.888184 & 1.012756 \\
\hline
13 & 0 & ---      & ---      & --- \\
   & 2 & 0.880049 & 1.018074 & 1.167549 \\
\hline
14 & 1 & 0.857394 & 0.995331 & 1.137504 \\
   & 3 & 0.898419 & 1.015636 & 1.173546 \\
\hline
15 & 0 & ---      & ---      & --- \\
   & 1 & 1.074658 & 1.444377 & 1.559163 \\
   & 3 & 0.832484 & 0.851653 & 1.01621  \\
\hline
18 & 1 & 0.778685 & 0.896073 & 1.021961 \\
   & 2 & 1.060496 & 1.22807  & 1.412578 \\
\hline
19 & 0 & ---      & ---      & --- \\
   & 1 & 1.002919 & 1.069563 & 1.226078 \\
   & 2 & 1.02404  & 1.191085 & 1.368696 \\
\hline
20 & 2 & 0.67185 & 0.805325 & 0.914874 \\
   & 3 & 1.07768 & 1.307121 & 1.502083 \\
\hline
21 & 0 & ---      & ---      & --- \\
   & 2 & 0.832815 & 0.947791 & 1.088174 \\
   & 3 & 1.17426  & 1.400983 & 1.619065 \\
\hline
26 & 1  & 0.781691 & 0.909825 & 1.036532 \\
   & 2  & 1.044972 & 1.217382 & 1.397128 \\
   & 3  & 1.01369  & 1.137329 & 1.319162 \\
\hline
27 & 0 & ---       & ---      & --- \\
   & 1  &\bf{1.0}  & 1.111333 & 1.276815 \\
   & 2  &\bf{1.0}  & 1.107638 & 1.270047 \\
   & 3  &\bf{1.0}  & 1.216526 & 1.408507 \\
\hline
\hline
\end{tabular}
    \caption{\label{tab3dd} Values of $\delta_i^{\star}$(s) for $p$ = 1, 2 and 3 and $d$ = 3, \sug{defined on the refined grid}. Bold values are for exact splitting.}
\end{table}

\begin{figure}[ht!]
\centering
\includegraphics[width=\textwidth]{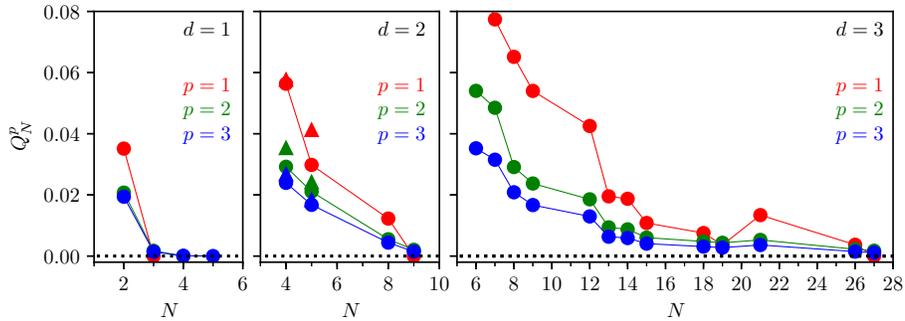}
    \caption{$Q_N^p$ values given by Eq. (\ref{eq-qdef}) as a function of $N$ for $p=1$ (red), $p=2$ (green) and $p=3$ (blue). Left panel is for $d=1$, middle panel is for $d=2$ and right panel is for $d=3$.}\label{fig-q}
\end{figure}

The most important for PIC simulations is to decrease as much as possible the value of $Q_N^p$. In Fig. \ref{fig-q} we show the optimal $Q_N^p$ as a function of the number of children $N$. For \sug{the 1D case ($d=1$, left panel)}, unsurprisingly, $Q_N^p$ decreases monotonically to zero: the exact solution is reached for $N=3$ with $p=1$, for $N=4$ with $p=2$ and for $N=5$ with $p=3$. Unless very strong constraints are placed on the number of children that can be used in a simulation, the exact solution can be used for $d=1$. For \sug{the 2D case ($d=2$, middle panel)}, things are quite similar: $Q_N^p$ decreases monotonically, and the exact solution is reached with $N=9$ for $p=1$. While for larger $p$ values the exact solution is not reached, the $Q_N^p$ decrease to values on the order of $10^{-3}$. One can also notice the gain obtained from $N=8$ to $N=9$, just by adding a child at the position of the parent.
 \sug{It is worth noticing that patterns of type 1 and 2 have the same number of children, namely 4. We then operate the optimization process for each cases ; the best values are depicted with solid circles in Fig. \ref{fig-q} while the worst ones are indicated with triangles. In order to identify the patterns of the worst cases (depending on the order $p$), the associated $w_j$'s and $\delta_j$'s in Tab. \ref{tab2dw} and \ref{tab2dd} are tagged with a dagger.}
A similar picture merges for the 3D case \sug{($d=3$, right panel)}. From $N=6$ to $N=27$, all $Q_N^p$ are decreasing with $N$, whatever the $p$ value, except for $N=21$. As for $d=1$ and $d=2$, the larger $p$, the smaller the $Q_N^p$. The values reached at $N=13$ are on the order of 5 $10^{-3}$ except for $p=1$ where such values of $Q_N^p$ are reached for $N=19$. The $Q_N^p$ would reach zero for the exact solution which needs far more children : $(p+2)^3$. We also computed for $N=12$ and $N=13$ the $Q_N^p$ values for patterns \{1,1\} and \{0,1,1\}. As for $d=2$, we obtained larger values, so these solutions are of no interest.

The parameters for exact splitting in two and three dimensions are not tabulated  (except for $d=2$ and $p=1$) because they involve some more complex patterns, that would need a different notation. But these values can be deduced very straightforwardly from the ones at one dimension from Eq. (\ref{eq-spk}).


\section{Density fluctuation level when splitting many particles}

While the values of $Q_N^p$ are important to evaluate the accuracy of the method, it is not that easy to interpret the associated consequences in a PIC code.
\sug{The level of density fluctuations is the root mean square of the density over a large enough number of grid points (for an acceptable statistics) in an homogeneous system. It has a strong impact on the stability of non-AMR PIC codes\cite{birdsall1991, winske2003} but in the AMR PIC context, a too large increase of the density fluctuation level can dramatically compromised the stability of such simulation\cite{fujimoto2011}. We hence evaluate how the density fluctuation level is modified by splitting a set of parent particles using the optimal splitting described above.}

\begin{figure}[ht!]
\centering
\includegraphics[width=\textwidth]{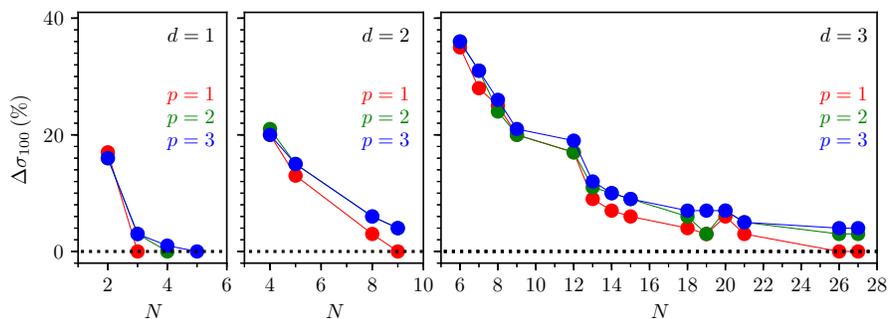}
\caption{\sug{Relative increase (in percentage) of the density fluctuation level for 100 particles per cell randomly and uniformly distributed in 4000 grid points} depending on $N$ for $p=1$ (red), $p=2$ (green) and $p=3$ (blue). Left panel is for $d=1$, middle panel is for $d=2$ and right panel is for $d=3$.}\label{fig-fluct}
\end{figure}

\sug{Whatever the dimension and the order of the B-splines, we dispatched 100 parent particles per cell over a grid of 4000 cells (plus the needed ghost particles on the border to avoid density drops). The value of 100 particles per cell is typical of PIC codes, and 4000 is large enough to carry statistical calculations (we obtained very similar curves using 2000 grid points). We then measured the 4000 density values at the center of each cells, in order to calculate $\sigma_{100}$ the standard deviation of this sample. This provides the level of fluctuation of the parents. Then, for all the $N$ values introduced in Tab. \ref{tab1dw} to \ref{tab3dd}, we split each parent particle, and calculate the associated density value at the same location, i.e. at the center of each cell. As for the parent, we then calculate over this sample of 4000 points the standard deviation $\sigma_{100}^{\star}$. Finally, Fig. \ref{fig-fluct} depicts the relative increase (in percentage) of the density fluctuation level.}

\sug{As a first remark, the level of fluctuation of the childern is always larger than the one of the parents. Without clear evidence, one could suspect that the larger total variation of the assignment function of the children compared to the one of the parent could play a significant role in this feature. Furthermore,}
Fig. \ref{fig-fluct} clearly shows that when increasing the number of children $N$, whatever the $d$ and $p$ values, the density level of fluctuations of the children gets closer to the one of the associated parents. This gives strength to this method: while increasing the number of children, the accuracy of the splitting is better, and the increase of the density level of fluctuations is lowered.

In Tab. \ref{tab2}, we computed the standard deviations for the increase of \sug{the density} fluctuation level for the Fujimoto (first line) Innocenti (second line) and this Work (third line) -type splitting, depending on the degree $p$ of the B-spline.
The calulations are conducted exactly as in the same way as described above to produce Fig. \ref{fig-fluct}.
\sug{In all cases, the children are dispatched using pattern 1, with $\delta_1^{\star}$ values equal 0.1, 0.25 and 0.955536 associated to Fujimoto, Innocenti and this work, respectively.}
These values are computed in the the 2D case for the B-spline degree $p=1$ (left row), $p=2$ (middle row) and $p=3$ (right row). The consequence of the splitting type on the level of fluctuation is clear and dramatic if using a non-optimized method. As already discussed, if these values happen to be unacceptable because too large, they can be decreased by adding some more patterns (and so the number of split particles) in the code provided.

\begin{table}
\centering
\begin{tabular}{l r r r}
\hline
\hline
Splitting type  & \hspace{0.0cm} $p=1$ & \hspace{0.0cm} $p=2$ & \hspace{0.0cm} $p=3$ \\
\hline
\hline
Fujimoto  & 122 \% & 116 \% & 114 \% \\
\hline
Innocenti & 106 \% & 105 \% & 106 \% \\
\hline
This work &  21 \% &  22 \% &  22 \% \\
\hline
\hline
\end{tabular}
\caption{\label{tab2} Mean values and standard deviations for the increase of fluctuation level for the Fujimoto (first line), Innocenti (second line), and Smets (third line) -type splitting depending on the $p$ degree of the spline. These values are obtained in the 2D case for a splitting using the four children of pattern 1.}
\end{table}

\section{Conclusions}

We presented a new method to find the optimum way to dispatch particles in PIC codes when a splitting strategy is needed in a AMR context. Splitting techniques have already been investigated and discussed, but not necessarily in the framework of AMR. In this specific case, the change of the grid mesh for the split particles has strong consequences. Contrary to most of the previous studies, we emphasized the fact that the way to dispatch split particles must depend on the degree $p$ of the B-spline used as assignment function for the particles. We also demonstrated that such splitting is optimum for a symmetrical distribution of the split particles, meaning that the number of split particles is constrained by this requirement. This symmetry is preserved by adding a particle in the set of split particles at the parent's particle locus. We then provide in this paper the loci of the split particles in Fig. \ref{fig-where}, and the associated weights and positions of these split particles in Tab. \ref{tab1dw} to \ref{tab3dd}, for first, second and third order of interpolation, and for both 1D, 2D and 3D cases. The Python code\cite{splitpic} we use for this optimization can then be run for refinement factor larger than two, not tabulated in this paper.

For the 1D cases, the exact solution is affordable as this type of simulations is generally light enough. For the 2D case, one already reaches a very good solutions for $N=9$. It means that the number of split particles is multiplied by 9, while the number of cells is multiplied by 4 (on a general point of view, it is multiplied by $r^d$). The number of particles per cells is then only multiplied by 2.25. For the 3D case, very good solutions are reached for $N=13$ ($p=1$ and $p=2$) or $N=19$ ($p=3$), while the number of cells is multiplied by 8. The number of particles per cell is then multiplied by 1.625 or 2.375. In any cases, the number of particles per cell is not prohibitive.

We emphasized the fact that the assignment function of the set of split particles has to be as close as possible to the assignment function of the particle to be split. This constraints is more important than the one generally considered\cite{assous2003},\cite{welch2007}, only associated to its integrated values: mass/charge, impulsion, energy. This point has already been pointed out\cite{pfeiffer2015} while conserving the full distribution function needs a heavier numerical effort. From a computational point of view, the method we present has the advantage of having absolutely no overhead associated to the calculations of the coefficients (weight and locus) if these values are calculated and tabulated. In previous studies, the splitting process can be aborted if at least one of the split particle is out from the cell (on the fine grid) of the parent locus. Such a constraint has the drawback to violate the symmetrical constraint on the number and loci of the children, discussed in this paper. This requirement seems to us unsuitable, and should be re-evaluated in the light of the method presented here.

We have also emphasized the importance for a set of split particle to have an associated assignment function as close as possible as the one of the particle which is split. In Maxwell's equations, these particles are playing a role through their zero order (charge density) and first order (current density) moment of their distribution function. The definition of these quantities being linear with the assignment function, the errors on this quantities will evolve in the same way as the one presented for the assignment function. This is clearly the case for the charge density, but also for the current density as all the split particles have the same velocity as the one of the particle that is split.

Up to now, all the calculations have been conducted with the assumption that the grid is isotropic, meaning that the mesh size is the same in all directions. Actually, it is not a constraint, and the above $w_i$ and $\delta_i$ values are also the optimum ones for anisotropic grid, whatever the degree of anisotropy. For the two imensional cases, a simple homothety centered on the parent position with a ratio $\Delta_Y / \Delta_X$ only in the $y$ direction is the link between the isotropic case with mesh size $\Delta_X$ and the anisotropic one with mesh size ($\Delta_X, \Delta_Y$). As this can be checked with the optimization code\cite{splitpic} (for both two and 3D cases), this transformation does not modify the optimization results, meaning that the optimum parameters are not depending on the degree of anisotropy of the grid.

These conclusions should also be considered in the opposite process of merging. Up to now, most of the existing methods identify a set of particles as close as possible in phase space, and merge two particles in a single one \cite{teunissen2014}, or a larger set of particles in a pair \cite{vranic2015}. In the merging strategy, the degree of the B-spline of the assignment function could also play an important role as in the splitting strategy, so the way to merge particles might also depend on this degree. To be more specific, the set of particles to be merged should be as close as possible in velocity space, but at an optimal finite distance in position space.

An important conclusion of this work is that when splitting a particle, the split particles should be dispatched at a given finite distance from the original particle, this distance depending only on the degree $p$ of the B-spline and on the dimension $d$ of the problem. We can draw an interesting parallel with a recent study on the merging problem by Luu et al. \cite{luu2016}. This study uses a Voronoi diagram to identify particles close enough in phase space to be merged. This algorithm needs a tolerance parameter, which is somewhat the threshold value below which one can consider that the particles to be merged are close enough. In this study by Luu et al., they show that this value has to be very small in velocity space, but can be much larger in position space. While not a proof, it suggests the importance of the finite distance between particles involved in merging or splitting processes.

We also outlined the existence of an exact splitting at the cost of a large number of split particles. Considering the existence of efficient rezoning algorithm\cite{pfeiffer2015}, one can wonder how the approximate splitting we present in this paper can compare with the exact one, followed by a rezoning procedure. Both options could be used in order to have the same number of split particles. The first option will be the cheapest in term of CPU, but their efficiency to preserve the distribution function and keep the fluctuations at the same level should be evaluated in a future work.

\textbf{\textsf{Acknowledgement}} : This work has been done within the LABEX Plas$@$par project, and received financial state aid managed by the Agence Nationale de la Recherche, as part of the programme ``Investissements d'avenir'' under the reference ANR-11-IDEX-0004-02.

\bibliographystyle{elsarticle-num}
\bibliography{cpc}

\end{document}